\documentclass[aps,prl,twocolumn,10pt,showpacs,superscriptaddress]{revtex4-1}
\usepackage[english]{babel}
\usepackage[utf8]{inputenc}
\usepackage{amsmath,amssymb,amsbsy,graphicx,braket,mathtools,wrapfig,hyperref,
color,epsfig,import,bm,soul}

\renewcommand{\v}[1]{\ensuremath{\mathbf{#1}}} 

\newcommand{\D}{^{\dag}}

\newcommand{\MD}{^{\mathstrut}}

\providecommand*{\iu}{\ensuremath{\mathrm{i}}}

\newcommand{\vssp}{\vspace{0.3cm}}

\begin{document}

\title{Topological Marker Currents in Chern Insulators}

\author{M. D. Caio}
\affiliation{Instituut-Lorentz, Universiteit Leiden, P.O. Box 9506, 2300 RA 
Leiden, The Netherlands} 
\author{G. M\"oller}
\affiliation{Functional Materials Group, School of Physical Sciences, 
  University of Kent, Kent CT2 7NZ, United Kingdom}
\author{N. R. Cooper}
\affiliation{T.C.M. Group, Cavendish Laboratory, J.J. Thomson Avenue,
  Cambridge CB3 0HE, United Kingdom}
\author{M. J. Bhaseen}
\affiliation{Department of Physics, King's College London, Strand,
  London WC2R 2LS, United Kingdom}

\begin{abstract}

Topological states of matter exhibit many novel properties due to the
presence of robust topological invariants such as the Chern index.
These global characteristics pertain to the system as a whole and are
not locally defined. However, local topological markers can
distinguish between topological phases, and they can vary in space. In
equilibrium, we show that the topological marker can be used to
extract the critical behavior of topological phase transitions. Out of
equilibrium, we show that the topological marker spreads via a flow of
currents, with a bounded maximum propagation speed. We discuss the
possibilities for measuring the topological marker and its flow in
experiment.

\end{abstract}

\maketitle

Topological quantum systems exhibit many striking phenomena due to the
inherent topological properties of their ground state wavefunctions.
Experimental signatures in two-dimensions include the robust
quantization of the transverse charge 
and spin
transport
, with direct links to topological
invariants~\cite{Thouless1998}
. The observation of the Quantum
Hall Effect in graphene highlights that topology can be relevant at
room temperature~\cite{Novoselov2007}, widening the scope for
practical applications. The recent discovery of
topological insulators~\cite{Bernevig2006,Konig2007,Fu2007} extends
the reach of topology to a wider class of materials and
dimensionalities, giving rise to exotic phases such as topological
superconductors~\cite{Sato2017}. Discoveries of
topological phases in photonic
systems~\cite{Carusotto2013,Lu2014,Fitzpatrick2016,LeHur2016,Noh2017} and cold atomic
gases~
\cite{Dalibard2011,Miyake2013,Esslinger2014,Goldman2014,Aidelsburger2014,Lohse2015,Schweizer2016,Goldman2016a,Goldman2016,Mukherjee2017}
have expanded the range of experimental probes and measurement
techniques,
providing access to a much broader range of physical observables.
These diverse systems could also play an important role in fault tolerant
quantum information processing~\cite{Nayak2008}.

Recently, the behavior of non-equilibrium topological systems has come
under scrutiny, with a view towards the time-dependent interrogation
and manipulation of their novel topological properties. Theoretical
studies include quantum quenches in $p+ip$
superfluids~\cite{Foster2013,Foster2014} and Chern
insulators~\cite{Caio2015,Caio2016}. Non-equilibrium dynamics of
topological systems has also been examined in the context of
periodically driven Floquet
systems~\cite{DAlessio2015,Wang2015,Wang2016,Dehghani2016,Yates2016},
as recently realized in experiment~\cite{Esslinger2014}. A notable
finding is that global topological invariants are preserved
under unitary
evolution~\cite{Foster2013,Foster2014,Caio2015,DAlessio2015},
unless dynamically-induced symmetry breaking
takes place~\cite{McGinley2018}. However, local physical observables, such as
the magnetization, can change~\cite{Caio2015}. In addition, the Hall
response is no longer quantized, and undergoes temporal
dynamics~\cite{Caio2016,Wang2015,Wang2016,Dehghani2016}.

In this work, we examine the equilibrium and non-equilibrium
properties of Chern insulators from the vantage point of the
real-space topological marker~\cite{Bianco2011}. We show that the
topological marker can be used to extract the critical
behavior of topological phase transitions,
in spite of the fact that no traditional local order parameter exists.
Out of
equilibrium, we show that the topological marker spreads via a flow of
currents, with a maximum propagation speed that is determined by the
band structure of the final Hamiltonian. We discuss the relevance of
these findings to experiment.

\begin{figure}
  \includegraphics[width=3.2in]{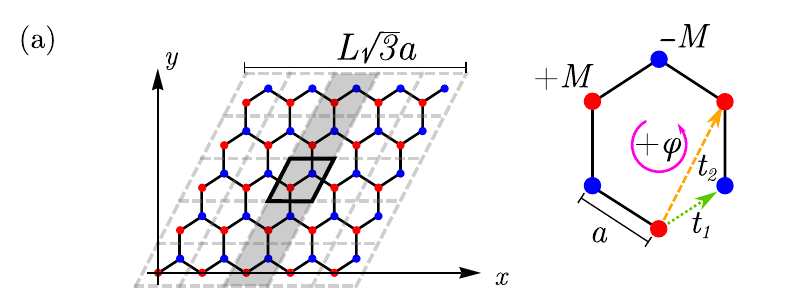}
  \includegraphics[width=3.2in,trim=0 5 0 8,clip]{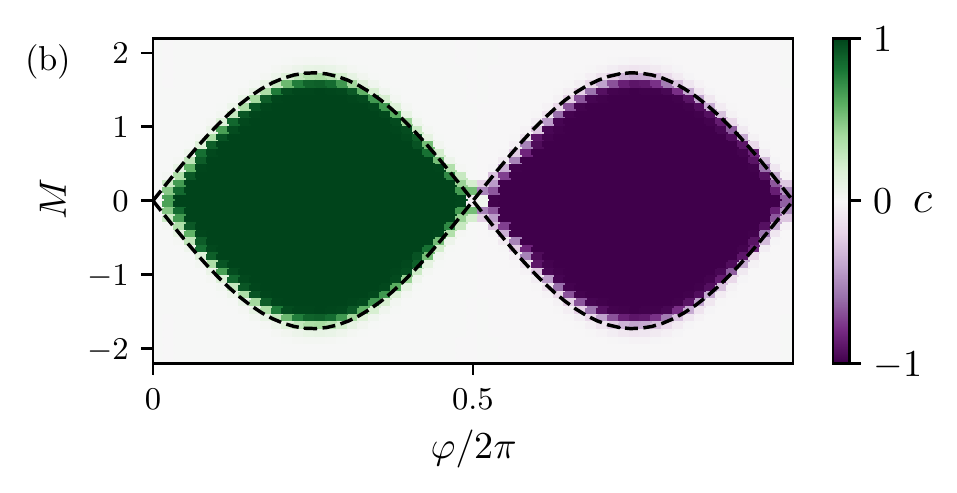}
  \caption{(a) Lattice geometry of the Haldane model. We consider a
    diamond-shaped sample (light grey) with edges composed of $L$ unit
    cells along primitive lattice vectors, and $N=2L^2$ sites. (b)
    Density plot of the Chern marker (\ref{chernmarker}) in the
    central cell (bold) in panel (a), with $L=17$. The dashed lines
    are the phase boundaries $M = \pm \sqrt{3}\sin \varphi$ of the
    Haldane model~\cite{Haldane1988} with $t_1=1$, $t_2=1/3$ and $a=1$.}
  \label{Fig:Phasediag}
\end{figure}
\vssp

{\em Model.}--- In order to expose the applications of the topological
marker, we focus on the Haldane model~\cite{Haldane1988}.  This model
is celebrated for realizing topological bands without an external
magnetic field, as recently exploited in cold atomic gas
experiments~\cite{Esslinger2014,Dalibard2011}. It also played an
  important role in the discovery of topological insulators~\cite{Kane2005}.
The Hamiltonian describes
spinless fermions on a honeycomb lattice
\begin{eqnarray}
 \hat H&=&-t_1\sum_{\langle i,j\rangle}\left(\hat c\D_i \hat 
c\MD_j+\mbox{h.c.}\right) - t_2 \sum_{\langle\!\langle 
i,j\rangle\!\rangle}\left(e^{\iu \varphi_{ij}} \hat c\D_i \hat 
c\MD_j+\mbox{h.c.}\right) \nonumber \\ & &+M \sum_{i\in A} \hat n_i -M 
\sum_{i\in B} \hat n_i,
\label{Haldane}
\end{eqnarray}
where the fermionic creation and annihilation operators $\hat
c_i^\dagger$ and $\hat c_i$ obey anticommutation relations $\{\hat
c_i,\hat c_j^\dagger\}=\delta_{ij}$, and $\hat n_i\equiv\hat
c_i^\dagger \hat c_i$; see Fig.~\ref{Fig:Phasediag}(a).  Here, $t_1$
and $t_2$ are the first and second neighbor hopping amplitudes, and
the angular brackets $\langle \rangle$ and $\langle\!\langle
\rangle\!\rangle$ indicate summation over the first and second
neighbor pairs respectively.  The parameter $M$ breaks the inversion
symmetry between the $A$ and $B$ sublattices, as indicated by red and blue dots in Fig.~\ref{Fig:Phasediag}(a), yielding a trivial
insulating phase for sufficiently large $M$.  The phase
$\varphi_{ij}=\pm \varphi$ breaks time-reversal symmetry, and is
positive (negative) for anticlockwise (clockwise) second neighbor
hopping, as shown in Fig.~\ref{Fig:Phasediag}(a). This allows for
topological phases at half-filling, even without a net magnetic
field~\cite{Haldane1988}. The phases are distinguished by the Chern
index, which is a global property of a band of Bloch states $|\psi(\v
k)\rangle$:
\begin{equation}
  C=\frac{1}{2\pi}\!\int_{\rm BZ} \!d^2k\,\,\Omega, \label{chernindex}
\end{equation}
where $\Omega=\partial_{k_x}A_{k_y}-\partial_{k_y}A_{k_x}$ is the
Berry curvature, $A_{k_\mu}=i\langle\psi(\v
k)|\partial_{k_\mu}|\psi(\v k)\rangle$ is the Berry connection, and
the integral is over the 2D Brillouin zone. The Chern index thereby
characterizes the topology of the band of states $|\psi(\v k)\rangle$
and is robustly quantized. For the ground state of the Haldane model
at half-filling, $C=\pm1$ in the topological phases and $C=0$ in the
non-topological phase; see Fig.~\ref{Fig:Phasediag}(b). Throughout the
manuscript, we fix the hoppings $t_1=1$, $t_2=1/3$, and the lattice
spacing $a=1$.

\vssp

{\em Local Chern Marker.}--- In open-boundary systems, or in the presence 
of disorder, the lack of translational invariance renders the expression
(\ref{chernindex}) undefined. Recently, the notion of a local Chern
marker has been introduced~\cite{Bianco2011}, with the explicit representation
\begin{equation}
 c({\bf r}_\alpha) = -\frac{4\pi}{A_c}{\rm Im} \sum_{s=A,B}\langle \v
 r_{\alpha_s}|\hat P \hat x \hat Q \hat y \hat P|\v
 r_{\alpha_s}\rangle,
  \label{chernmarker}
\end{equation}
where $A_c$ is the area of a real-space unit cell, $\hat P$ is the
projector onto the ground state, $\hat Q = \hat I - \hat P$ is the
complementary projector, and the sum is over the two sublattice sites
within the unit cell $\alpha$. Owing to the shortsightedness of $\hat
P$ for gapped phases~\cite{Kohn1996}, the
topological marker defined in Eq.~(\ref{chernmarker}) has a
quasi-local character~\cite{Resta2006,Resta2011}.
In particular, away from the phase boundaries,
in gapped phases, $\hat P$ is exponentially localized.  In order
  to orient the subsequent discussion, in Fig.~\ref{Fig:Phasediag}(b), we show
the phase diagram of the Haldane model (\ref{Haldane}) obtained from
the real-space Chern marker (\ref{chernmarker}); see also Ref.~\cite{Bianco2011}.
The Chern marker clearly discriminates between the topological and
non-topological phases~\cite{Bianco2011}, in accordance with the phase
diagram obtained from the low-energy Dirac theory~\cite{Haldane1988}.
In finite-size samples, the Chern marker averages to zero, but in the
interior of the sample it nonetheless distinguishes between
topological and non-topological phases~\cite{Bianco2011}. For
  recent applications of the Chern marker in clean and disordered
  samples see Refs.~\cite{Tran2015,Privitera2016,Tran2017,Marrazzo2017,Amaricci2017}.

\vssp

\begin{figure}
  \includegraphics[width=3.2in]{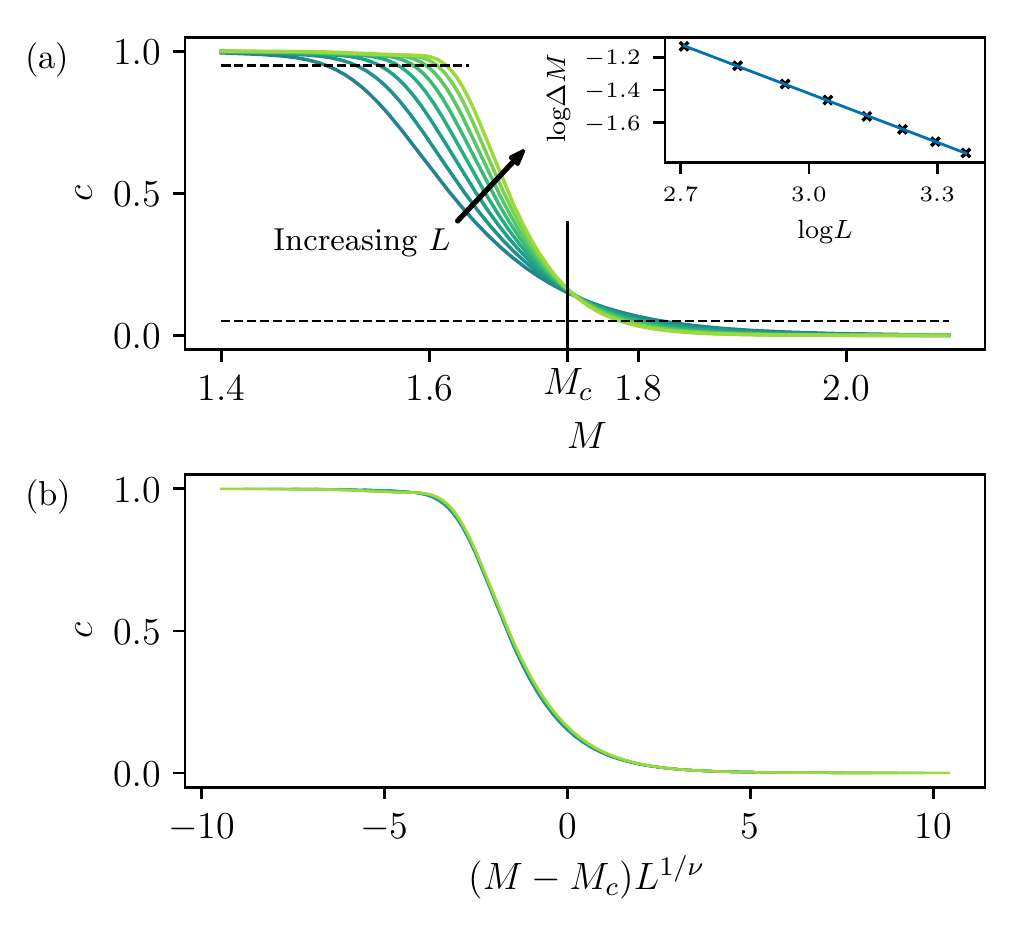}
  \caption{(a) Vertical slice through the phase diagram in
    Fig.~\ref{Fig:Phasediag}(b) with $\varphi=\pi/2$, showing the
    variation of the local Chern marker in the center of the sample as
    a function of $M$. The results smoothly interpolate between $1$
    and $0$ in the vicinity of the topological phase transition. The
    different curves correspond to increasing system sizes
    $L=15,17,19,21,23,25,27,29$, as illustrated. The curves cross in a
    narrow region within $0.5\%$ of $M_c=\sqrt{3}$, the exact
    transition point of the Haldane model for the chosen
    parameters. The width $\Delta M$ of the transition region,
    corresponding to the interval from $c=0.95$ to $c=0.05$
    (horizontal lines), scales as $\Delta M \sim L^{-1/\nu}$. Inset:
    linear plot showing $\nu = 0.995(6)\approx 1$. This is consistent
    with the correlation length exponent in the low-energy Dirac
    theory. (b) Re-plotting the data in panel (a) as $c\sim \tilde
    f((M-M_c)L^{1/\nu})$ with $\nu=1$ yields scaling collapse.}
  \label{Fig:Chernmarker}
\end{figure}

{\em Critical Properties.}--- Inspection of the phase diagram in
Fig.~\ref{Fig:Phasediag}(b) highlights that, away from the phase
boundaries, the local Chern marker is $0,\pm 1$ within machine
precision. However, in the vicinity of the transition for finite-size
samples, $c$ is no longer quantized~\cite{Bianco2011}. In
Fig.~\ref{Fig:Chernmarker}(a), we show the variation of the Chern
marker as one passes between the topological and non-topological
phases.
It can be seen that the transition region narrows with increasing
system size, suggesting a sharp discontinuity in the thermodynamic
limit. Assuming that the departure from quantization in the middle of
the sample occurs when the correlation length $\xi$ is of order (half
the) system size, finite-size effects become relevant when $\xi\sim
L/2$. Further assuming that $\xi \sim (\Delta M)^{-\nu}$, where
$\Delta M$ is the width of the transition region in
Fig.~\ref{Fig:Chernmarker}(a) and $\nu$ is the correlation length
exponent, one expects that the width scales with the system size
according to $\Delta M \sim L^{-1/\nu}$. In the inset of
Fig.~\ref{Fig:Chernmarker}(a) we confirm this dependence, with
$\nu=0.995(6) \approx 1$. This is consistent with the correlation
length exponent of the low-energy Dirac theory. It is also compatible
with the delocalization of the edge states into the interior of the
sample on closing the gap.  Re-plotting the data in
Fig.~\ref{Fig:Chernmarker}(a) with the scaling form $c\sim
f(\xi/L)=f((M-M_c)^{-\nu}/L)=\tilde f((M-M_c)L^{1/\nu})$ with $\nu=1$,
shows that the data collapse onto a single curve; see
Fig.~\ref{Fig:Chernmarker}(b). This confirms that the real-space
  Chern marker can be used to extract the critical behavior of
  topological phase transitions, in a similar way to a local order
  parameter for conventional phase transitions.  In contrast to
  approaches using the momentum space Berry curvature~\cite{Chen2017,Kourtis2017}, the present
  technique can be applied in non-translationally invariant
  settings.

\vssp

\begin{figure}
  \includegraphics[width=3.2in,clip]{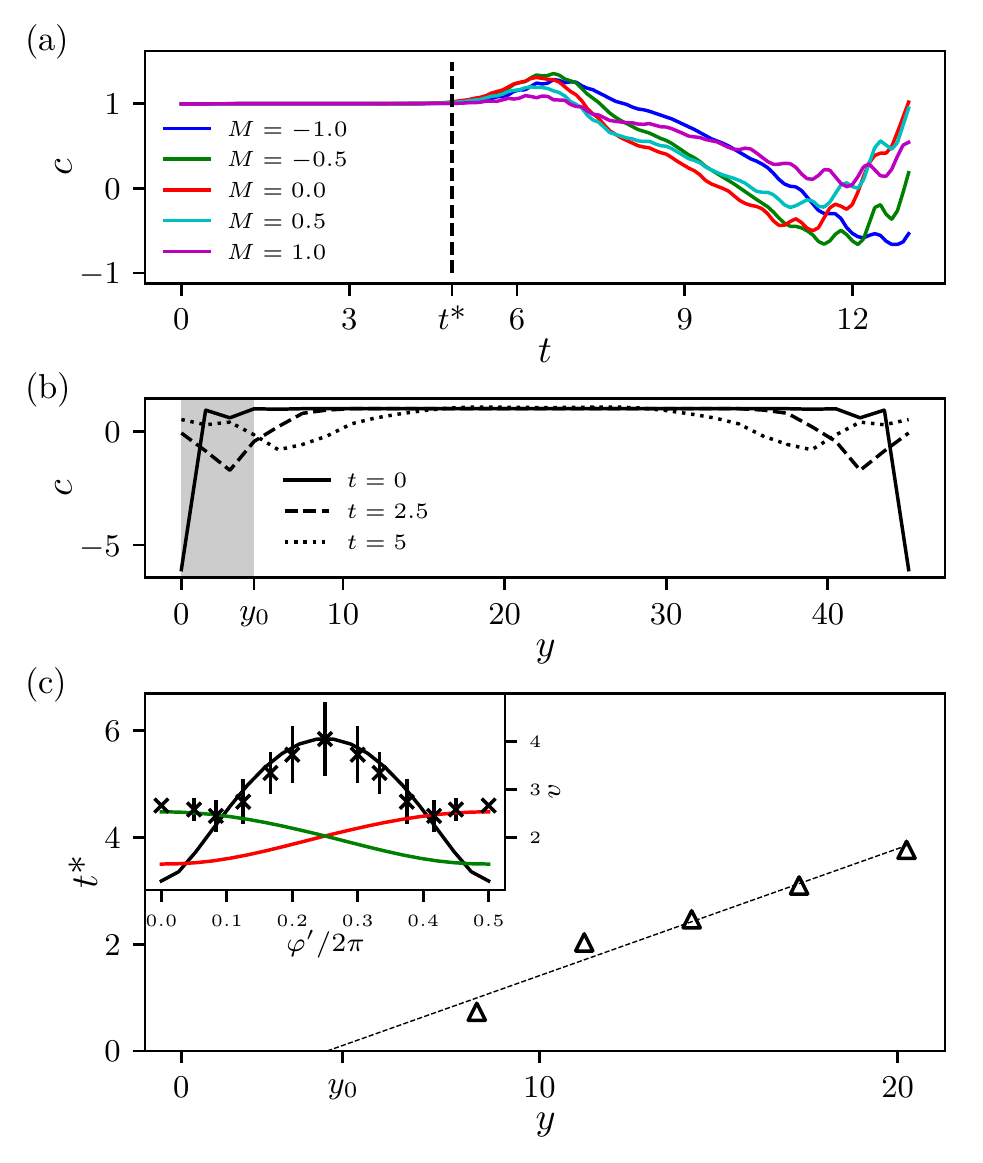}
  \caption{(a) Time-evolution of the Chern marker in the center of a
    sample with $L=31$, following quenches from different points in
    the topological phase with $M=-1,-0.5,0,0.5,1$, to the
    non-topological phase with $M^\prime=5$, and $\varphi=\pi/2$ held
    fixed. The Chern marker remains quantized until a characteristic
    time $t^*$, which is independent of the initial parameters. (b)
    Spatial profile of $c(\v r)$ along a cut corresponding to the
    shaded area in Fig.~\ref{Fig:Phasediag}(a), where $y$ is the
    distance from the boundary. The results are shown at times $t=0$
    (solid), $2.5$ (dashed), $5$ (dotted), following a quench from
    $M=0$ to $M'=5$, with $\varphi=\pi/2$ held fixed. At $t=0$, the
    edge has a width $y_0 \sim 4.5$ (shaded). As $t$ increases, a
    wave-like disturbance in $c(\v r)$ propagates into the
    interior. (c) Dependence of $t^\ast$ on $y$, for different system
    sizes $L=11,15,19,23,27$, for the quench considered in (b). A
    linear fit yields a propagation speed $v\sim 4.06\pm 0.77$; the
    $y$-intercept $y\sim 4.4\pm 2.0$, is close to the initial width of
    the edge. Inset: variation of $v$ with the parameter
    $\varphi^\prime$ of the final Hamiltonian, with $M^\prime=5$ and
    initial parameters $\varphi=\pi/2$ and $M=0$. The 
    speed $v$ corresponds to the maximum speed permitted by the final
    band structure. It coincides with the maximum of $v_1^y(\v k)=\partial
    E_1(\v k)/\partial k_y$ (red), $v_2^y(\v k)=\partial E_2(\v k)/\partial k_y$ (green)
    or $|v_2^y(\v k)-v_1^y(\v k)|$ (black), extremized over $\v k$-space, where
    $E_1(\v k)$ and $E_2(\v k)$ are the energies of the lower and
    upper bands respectively; see text.}
  \label{Fig:Noneq}
\end{figure}

{\em Non-Equilibrium Dynamics.}--- Having exposed the equilibrium
properties of the Chern marker, we turn our attention to its
non-equilibrium dynamics. Here, we focus on quantum quenches, where
the system is prepared in the ground state $|\psi_0(M,\varphi)\rangle$
of the initial Hamiltonian $\hat H(M,\varphi)$ at half-filling and,
upon sudden change of the parameters to new values
$(M^\prime,\varphi^\prime)$, it evolves as $\exp{[-i\hat
    H(M^\prime,\varphi^\prime)t]}|\psi_0(M,\varphi)\rangle$. The
  Chern marker is evaluated using Eq.~(\ref{chernmarker}), and the
  projector onto the time evolving state~\cite{Privitera2016}.  In
Fig.~\ref{Fig:Noneq}(a), we show the dynamics of the Chern marker,
evaluated in the center of a finite-size sample, following quenches
for different starting points in the topological phase to a fixed
parameter point in the non-topological phase.
For these quenches, the Chern marker initially remains quantized, but
it departs from quantization after a characteristic timescale
$t^\ast$, that is independent of the initial Hamiltonian
parameters. As we shall discuss in more detail below, this timescale
is consistent with a flow of Chern marker currents, from the vicinity
of the sample boundaries towards the interior. Evidence for the
propagation of these currents can be seen in real-space plots of $c(\v
r)$ along a cut through the center of the sample, as indicated by the
grey shaded region in Fig.~\ref{Fig:Phasediag}(a). The Chern marker,
which is initially negative near the sample boundaries (to ensure that
$c(\v r)$ integrates to zero) flows from the
edges towards the interior; see Fig.~\ref{Fig:Noneq}(b). Probing the
time at which $c$ departs from unity, in the middle of a sample for
different system sizes $L$, allows us to estimate the speed of
propagation, as illustrated in Fig.~\ref{Fig:Noneq}(c). In addition to
a well-defined propagation speed $v$, it is evident that the
disturbance emanates from the vicinity of the edges, where $y_0$ is
the finite width of the edge at equilibrium; see
Fig.~\ref{Fig:Noneq}(b).

{\em Propagation Speed.}--- In the inset of Fig.~\ref{Fig:Noneq}(c) we
show the non-trivial variation of $v$ for quenches to different points
in the phase diagram. It can be seen that $v$ coincides with the
maximum speed (in this case in the $y$-direction) allowed by the band
structure, where we measure speeds in units of $at_1/\hbar$.
Depending on the final parameters this is either the
maximum speed permitted by the upper and lower bands, or the maximum
of the relative band velocities, extremized over ${\bf k}$-space. The
latter is attributed to coherent particle-hole excitations following
the quench, and the presence of the excited state projector
$\hat Q=\hat I-\hat P$ in the definition (\ref{chernmarker}),
yielding interference terms
oscillating at the frequency of the band gap $\Delta(\v k)=E_2(\v
k)-E_1(\v k)$. The associated propagation speed of the Chern marker
can thus be larger than the individual band speeds (e.g. if the bands have slopes with opposite signs) as shown in
the inset of Fig.~\ref{Fig:Noneq}(c).

\vssp

{\em Topological Marker Currents.}--- Evidence for the propagation of
Chern marker currents can also be obtained from the dynamics of $c(\v
r)$ close to the sample edges. For a finite-size sample with open
boundaries, $\int c(\v r) d^2r=0$ at all times.  A local Chern current $\v
J_c$ therefore exists, such that $ \frac{\partial c}{\partial
  t}+\nabla\cdot {\v J}_c=0 $. In integral form, the flux of the Chern
current ${\mathcal F}_{\rm c}$ out of a unit cell is given by
\begin{equation}
 {\mathcal F}_{\rm c}:=\oint_{\partial A_c} \v J_c 
\cdot d\v l=-\int_{A_c}\frac{\partial c}{\partial t}\, d^2r, 
\label{Continuity}
\end{equation}
where $A_c$ and $\partial A_c$ are the area and perimeter of a unit cell. 
\begin{figure}
  \includegraphics[width=3.2in,clip]{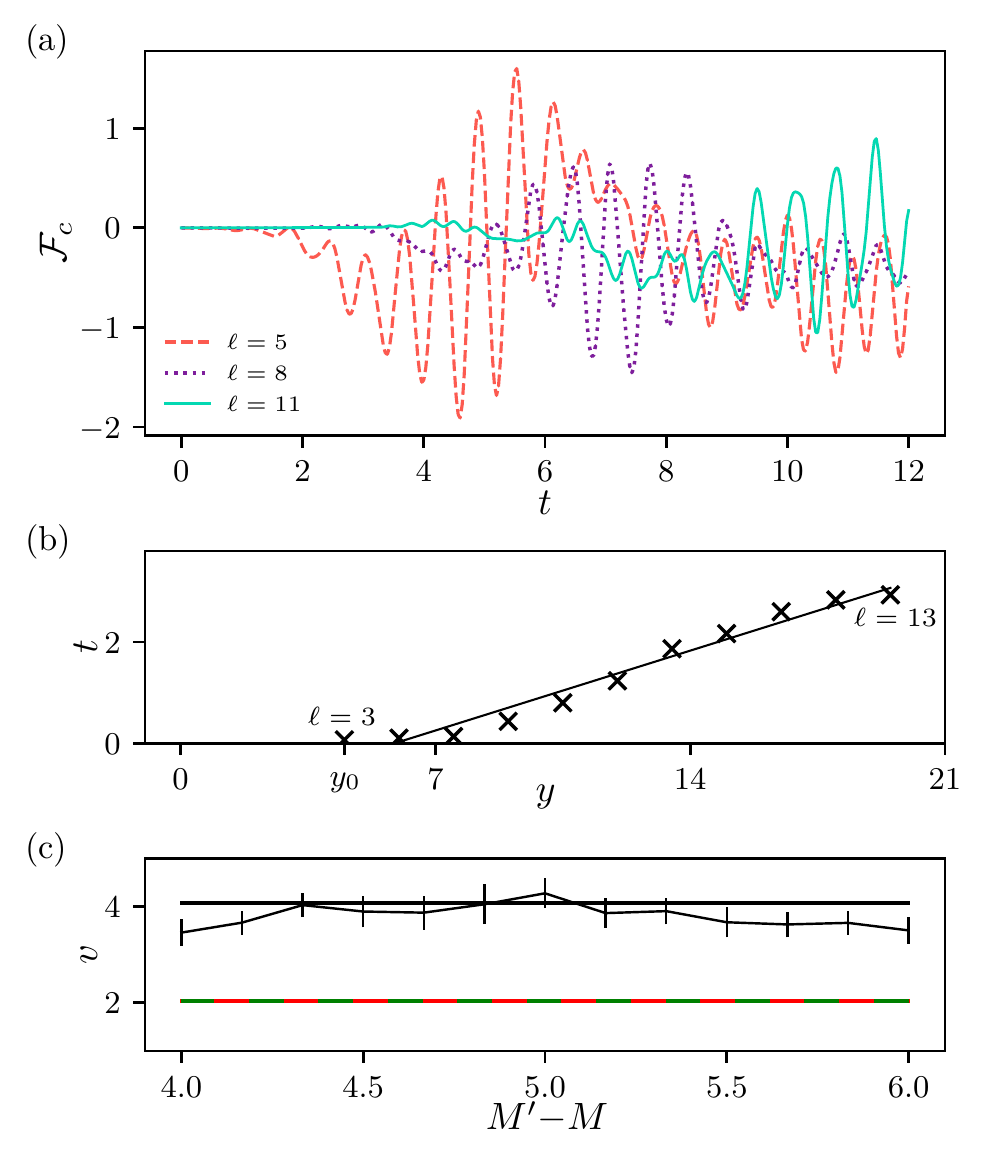}
  \caption{ (a) Flux of the Chern current ${\mathcal F}_{\rm c}$
    through a unit cell distant $\ell = 5$ (dashed), $8$ (dotted), and
    $11$ (solid) unit cells from the boundary of a sample with $L=31$,
    following a quench from $M=0$ to $M'=5$, with $\varphi=\pi/2$ held
    fixed. The flux departs from zero at later times as $\ell$
    increases, corresponding to the propagation of Chern currents from
    the edges.  (b) Onset time of ${\mathcal F}_{\rm c}\neq 0$ versus
    $y=3/2\, \ell$, for $\ell=3,4,5,\dots 13$. A linear fit yields a
    speed $v$, corresponding to the maximum speed allowed by the final
    band structure. The $y$-intercept is compatible with the initial
    edge width, $y_0$. (c) Variation of the speed $v$ for quenches
    from different points in the phase diagram with $M\in [-1,1]$ to
    $M^\prime =5$, with $\varphi=\pi/2$ held fixed. The speed
    is consistent with the maximum of $|v_2^y(\v k)-v_1^y(\v k)|$
    allowed by the final band structure (solid line), and exceeds the
    maximum speed of each band separately (red and green lines).}
  \label{Fig:Topcurrents}
\end{figure}
In Fig.~\ref{Fig:Topcurrents}(a), we plot the right hand side of
Eq.~(\ref{Continuity}), corresponding to the integrated flux of $\v
J_c$ through the perimeter of a unit cell, at different spatial
positions along a cut through the sample, as illustrated in
  Fig.~\ref{Fig:Phasediag}(a). The onset of a
non-vanishing flux ${\mathcal F}_{\rm c}$ occurs at later times with
increasing distance from the boundaries. The associated propagation
speed $v$ can be extracted from a linear fit of the onset time versus
distance, as shown in Fig.~\ref{Fig:Topcurrents}(b). In
Fig.~\ref{Fig:Topcurrents}(c) we plot $v$ as a function of the initial
parameters for a fixed final Hamiltonian. The extracted speed is approximately
independent of the initial starting parameters. In this particular case,
the speed $v$ is consistent with the maximum value of $|v_2^y(\v
k)-v_1^y(\v k)|$ for the final band structure, which exceeds the
maximum speed of each band separately.

\vssp

{\em Experiment.}--- Although the definition of the Chern marker
(\ref{chernindex}) may appear complicated, its static and dynamic
properties could be accessible in experiment. This could be done via
measurements of the projection operator $\hat P$, as recently
performed in photonic topological systems in a real-space
  basis~\cite{Schine2018}. The projector could also be measured using
quantum gas microscopes~\cite{Bakr2009}, based on recent proposals to
extract the single-particle density matrix~\cite{Ardila2018}.
Explicitly, this can be seen by inserting complete set of states
$\sum_{\gamma,s} |r_{\gamma_s}\rangle\langle r_{\gamma_s}|={\hat I}$
into Eq.~(\ref{chernmarker}), and noting that the matrix elements of
the projector $P_{\alpha_s \beta_{s^\prime}}=\langle r_{\alpha_s}|\hat
P|r_{\beta_{s^\prime}}\rangle=\sum_{E_k<E_F} \langle
r_{\alpha_s}|\psi_k\rangle \langle\psi_k|r_{\beta_{s^\prime}}\rangle$
are those of the single-particle density matrix. The evaluation of
$c(\v r_\alpha)$ follows, as $|r_{\gamma_s}\rangle$ is a natural basis
for the operators $\hat x$ and $\hat y$.  In equilibrium, the Chern
marker is also related to the local magnetization~\cite{Bianco2013},
allowing further possibilities for experimental investigation
~\cite{Nathan2017}.

\vssp

{\em Conclusions.}--- In this work we have examined the equilibrium
and non-equilibrium properties of the real-space Chern
marker. In equilibrium, we have shown that it can be used to extract
the critical properties of topological phase transitions, in a
  similar way to a local order parameter for conventional
  transitions. Out of equilibrium, $c(\v r)$ undergoes dynamics,
giving rise to a flow of topological marker currents with a bounded
propagation speed.
There are many directions for theory and experiment, including the
impact of disorder and interactions on the flow of topological marker
currents, and their realizations in other settings.  It would
  also be interesting to explore the possibilities for manipulating
  these currents.

\vssp

{\em Acknowledgments.}--- We are grateful for helpful conversations
with Hannah Price, Lorenzo Privitera, Vincent Sacksteder and Michel
Fruchart. This research was supported by the Netherlands Organization
for Scientific Research (NWO/OCW), an ERC Synergy Grant, EPSRC Grants
EP/K030094/1 and EP/P009565/1, the Simons Foundation, and the Royal
Society Grant No. UF120157. MDC and MJB thank the Thomas Young Centre
and the Centre for Non-Equilibrium Science (CNES) at King's College
London.

\end{document}